\documentclass[fleqn,usenatbib]{mnras}

\usepackage{newtxtext,newtxmath}
\usepackage[T1]{fontenc}
\usepackage{ae,aecompl}

\usepackage{graphicx}	% Including figure files
\usepackage{amsmath}	% Advanced maths commands
\usepackage{amssymb}	% Extra maths symbols

\newcommand{\myemail}{sgao@bnu.edu.cn}
\newcommand{\fb}{f_{\rm b}}
\newcommand{\sig}{\sigma_{RV}}
\newcommand{\dv}{\Delta{v_r}}

\newcommand{\Msun}{M_\odot}
\newcommand{\teff}{T_{\rm eff}}
\newcommand{\feh}{\rm{[Fe/H]}}

\title[The binarity of Galactic dwarf stars]{The binarity of Galactic dwarf stars along with effective temperature and metallicity}

\author[S. Gao]{
Shuang Gao$^{1}$\thanks{E-mail: \myemail}, He Zhao$^{1}$, Hang Yang$^{1}$, Ran Gao$^{1}$
\\
$^{1}$Department of Astronomy, Beijing Normal University, Beijing 100875, China\\
}

\date{Accepted XXX. Received YYY; in original form ZZZ}

\pubyear{2016}

\begin{document}
\label{firstpage}
\pagerange{\pageref{firstpage}--\pageref{lastpage}}
\maketitle

\begin{abstract}
The fraction of binary stars ($\fb$) is one of most valuable tool to probe the star formation and evolution of multiple systems in the Galaxy. We focus on the relationship between $\fb$ and stellar metallicity ($\feh$) by employing the differential radial velocity (DRV) method and the large sample observed by the Large Sky Area Multi-Object Fiber Spectroscopic Telescope (LAMOST). Main-sequence stars from A- to K-types in the third data release (DR3) of LAMOST are selected to estimate $\fb$. Contributions to a profile of DRV from radial velocity (RV) error of single stars ($\sig$) and orbital motion of binary stars are evaluated from the profile of DRV. Finally, we employ 365,911 stars with randomly repeating spectral observations to present a detailed analysis of $\fb$ and $\sig$ in the two-dimensional (2D) space of $\teff$ and $\feh$. The A-type stars are more likely to be companions in binary star systems than other stars. Furthermore, the reverse correlation between $\fb$ and $\feh$ can be shown statistically, which suggests that $\fb$ is a joint function of $\teff$ and $\feh$. At the same time, $\sig$ of the sample for different $\teff$ and $\feh$ are fitted. Metal-rich cold stars in our sample have the best RV measurement.
\end{abstract}


%%%%%%%%%%%%%%%%%%%%%%%%%%%%%%%%%%%%%%%%%%%%%%%%%%

%%%%%%%%%%%%%%%%% BODY OF PAPER %%%%%%%%%%%%%%%%%%

\section{Introduction}

Binary stars (and multiple systems) in the Milky Way play a significant role from stellar evolution to supernova \citep{Duchene2013}. However it is very challenging to identify a binary system from single stars and to determine physical parameters of the binary stars (mass ratios\footnote{The $q$ is defined as the ratio of the secondary to primary mass, i.e. $q\equiv M_2/M_1$} $q$, orbital periods $P$, and eccentricities $e$) in different stellar populations and Galactic environments.

The $\fb$ in different environments and types were investigated in the last decades. \cite{Duquennoy1991} and \cite{Raghavan2010} found that $\fb$ for solar-type stars is about $45$ per cent generally. More particularly, they reported that the $\fb$ increases from $40$ per cent for sub-solar to $60$ per cent for super-solar stars. For stars with lower masses ($<0.5 \,\Msun$), \cite{Dieterich2012} reported the $\fb$ is only about $26$ per cent. On the other hand, OB stars likely have much higher $\fb$. The $\fb$ of stars with $M>5 \,\Msun$ is larger than $70$ per cent \citep{Sana2012}. Many above studies have been conducted the $\fb$ as a steep and monotonic function of stellar mass. For selected MS stars, the relationship can be considered as a function of the $\fb$ and stellar effective temperature $\teff$.

% binary fraction
The radial velocity (RV, $v_r$) of the star can be determined by running a pipeline of spectra data. A single star has the constant RV with observed uncertainty $\sig$, meanwhile binary star systems show systematic shifts in RV caused by the orbital motion of each companion. RV changes help distinct binaries from single stars. The differential RV (DRV, $\dv$) method makes use of twice RV measurements to obtain the profile of DRV, that is the sum of two components of single and binary stars.

% binary fraction with metallicity
Recently, several investigators have studied $\fb$ that suffers from stellar metallicity. \cite{Gao2014} reported that the metal-poor stars are likely to possess a close binary companion than metal-rich population. The $\fb$ in Galactic halo and thick disk are $20$ to $30$ per cent larger than thin disk for close binaries with $P<1000\, \rm{d}$. This general trend was supported by \cite{Yuan2015} by stellar locus outlier method. Due to limitations of the number of sources and time spans, the $\fb$ estimated by these work have relatively large uncertainties, and $\fb$ of the Galactic thick disk by \cite{Yuan2015} is inconsistent with \cite{Gao2014}'s result. Besides, results of $\fb$ from these work are dominated by wider-period binaries. \cite{Hettinger2015}, however, revealed the multiplicity of short-period binaries with the sub-exposures data archived in the Sloan Digital Sky Survey (SDSS). The investigation found that metal-rich disk stars are $30$ per cent more likely to have companions with $P<12\,\rm{d}$ than metal-poor halo stars. Therefore the relationship between $\fb$ and stellar metallicity ($\feh$) is still too far to be solved entirely.

% method description
Large spectral surveys provide opportunities to obtain physical parameters, such as RV, of a massive amount of stars. To access $\fb$ by using DRV method, large sample with repeat RV observations during different epochs is necessary. 

% To fit binary component, we need to build the RV model based on the Keplerian orbital motion. The simple two-body motion follows a stable Keplerian orbit when the distance between two centres of masses is much larger than the radius of each object. For our case, binary stars with periods between a few days and years satisfy above assumption. Based on LAMOST spectral data, we apply DRV method to fit the contribution fraction $\fb$ of the binary component and observed error $\sig$ of stellar RV. This process is similar to the procedure adopted by \cite{Gao2014}. We try to estimate $\fb$ and $\sig$ along with different $\feh$. The two-dimensional space of $\teff$ and $\feh$ is employed to analyse definite trend of the $\fb$ with stellar characters. 

LAMOST sample and data reduction are described in Section \ref{sec:data}. Our analysis method is presented in Section \ref{sec:method}. The results are presented and discussed in Section \ref{sec:result}. A conclusion is given in Section \ref{sec:summary}.

\section{Data}
\label{sec:data}
The Large Sky Area Multi-Object Fiber Spectroscopic Telescope \citep[also known as LAMOST, see][for the overview]{Cui2012,Zhao2012} is an optical telescope operated by National Astronomical Observatories of China (NAOC). It is special reflecting Schmidt system with 4-meter diameter and 4000 fibres in a large field of view (FOV) of 20 deg$^2$ in the sky. The LAMOST Experiment for Galactic Understanding and Exploration (LEGUE) survey of Milky Way stellar structure observed spectra of stars and released the third data (DR3) recently\footnote{ http://dr3.lamost.org}, containing 5,755,126 spectra and 3,178,057 \emph{AFGK} stars with stellar parameters in DR3 catalogue totally. Those stars were observed from 2011 October 24 to 2015 May 30.

DR3 catalogue gives the stellar physical parameters $\teff$, $\feh$, and surface gravity ($\log{g}$) of 3,178,057 A, F, G and K-type stars based on template match method \citep{Wu2011,Wu2014}. The determination of these parameters consistently adopts the cross-correlation approach with a simultaneous linear combination of single stellar templates \citep{Luo2012}.

We identify 365,911 stars that were observed twice from almost 3.2 million sources. The RV of two epochs of each star can be obtained via self-matching of the catalogue. Among these stars, our sample is selected by using the following criteria:

\begin{itemize}
  \item  $\rm{SNR}g > 20 $;
  \item  $-300\, \rm{km\,s^{-1}}< \dv < 300 \,\rm{km\,s^{-1}}$;
  \item  $4000 \,\rm{K} < \teff < 8000 \,\rm{K}$;
  \item  $-1.0\,\rm{dex} < \feh < 0.5\,\rm{dex}$.
\end{itemize}

Where $\rm{SNR}g$ is signal-noise-ratio at $g$ band and $\dv$ is defined as the difference between two RV observations $v_1-v_2$ to select MS dwarf stars, the $\log{g}$ is constrained sectionally following \cite{Liu2014}:

\begin{itemize}
  \item $\log{g} > 4.0$, if $\teff < 5500\, \rm{K}$;
  \item $\log{g} > 3.5$, if $5500 \,\rm{K} < \teff < 6000 \,\rm{K}$;
  \item $\log{g} > 3.0$, if $ \teff > 6000 \,\rm{K}$.
\end{itemize}

Further, the sources observed on the same nights are excluded because so small time span cannot generate significant enough RV shift of binary stars. Besides, observations in the same nights may underestimate random errors of RVs. Finally, our sample contains 303,803 quality dwarf stars with their parameters pairs from repeated observations. 

\begin{figure}
	\includegraphics[width=\columnwidth]{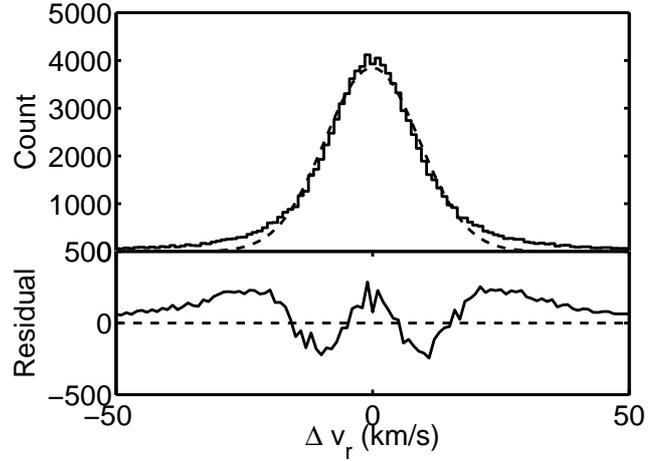}
    \caption{The profile of DRV and fitting. \emph{Top}: The distribution of $\dv$ of the entire sample. The solid and dashed curve is the histogram of observed data and a Gaussian fitting result, respectively. \emph{Bottom}: The residual error of data and Gaussian fitting. The x-axis is limited within $-50$ to $50\,\rm{km\,s^{-1}}$ to obtain a more distinct presentation. The bin size of x-axis is $1\,\rm{km\,s^{-1}}$.}
    \label{fig:dvall}
\end{figure}

The general distribution of $\dv$ shows a non-Gaussian symmetric profile as Figure \ref{fig:dvall}. One Gaussian function cannot fit our data even though the profile of $\dv$ is similar with a bell-shaped curve. The significant residual errors between data and one Gaussian fit imply that only random error caused by single stars observation is inadequate to explain $\dv$ data.

\textbf{The pipelines of LAMOST have a validation with SDSS data. The stellar parameters in our sample have been determined by an automatic process based on \citealt[]['s]{Wu2011} method. Gao et al. (2015) found that the standard deviations of $110\,\rm{K}$, $0.11\,\rm{dex}$ and $4.91\,\rm{km\,s^{-1}}$, for $\teff$,  $\feh$ and RV. According to the work of \cite{Ren2016}, LAMOST's $\teff$ and $\feh$ have the deviation of about $100\,\rm{K}$ and $0.1\,\rm{dex}$ in Kepler field, respectively. The pipeline and these independent validations confirm the accuracy of $\teff$ and $\feh$ for dwarfs, which is matching the coverage of parameters in our sample. Our analysis is based on the sample and the errors of stellar parameters. The unprecedented size of the sample with reliable errors help distinguish binary stars from single stars and draw the potential trend between the binarity and stellar parameters.}

% 3
\section{Method and Model}
\label{sec:method}

The spectral observations of two epochs may obtain different $v_r$ of a star. These mechanisms work to a shift of $v_r$: pulsating star, observed random error, and orbital motion of the binary system. We ignore the first one for MS stars in our mass range because the tiny deviation of MS stars cannot be uncovered by an observed accuracy of $\rm{km\,s^{-1}}$-size. The latter two mechanisms change observed RV of stars at two epochs with the different way: all RV data suffer observed random error but orbital motion only affects a fraction of stars, that is our scientific goal $\fb$. We develop a simple method to estimate respective contributions of the last two mechanisms to some given sample without known $\fb$ and $\sig$.

A sample with a mixture of single and binary stars from observed data, like LAMOST spectral database, needs to be revealed the ratio of the contributions caused by two components. Pure single stars and binary stars components are derived via simulations of our model that involved random error and Kepler's orbital motion.

% 3.1
\subsection{Model and Simulation}

For a single star, observed $v_r$ suffers random error $\sig$. We assume that the $\dv$ between two epochs follows a Gaussian distribution with an average value of 0 and standard deviation of $\sqrt{2}\sig$:
\begin{equation}\label{eq:single}
  \dv^{(s)}=v_1^{(s)}-v_2^{(s)} \sim N(0,\sqrt{2}\sig),
\end{equation}
where superscript $(s)$ represents single star and $(b)$ in the following text is for binary stars.

For a binary star system, besides random error, observed $v_r$ changes with the orbital motion of companions. The profile of $\dv^{(b)}$ is so complicated that we have to describe it by using a model that involves $P$, $q$, the mass of primary star $M_1$ and other orbital parameters.

Given $P$, $M_1$ and $q$, the maximum shift $K$ of RV of the primary star can be calculated based on the Keplerian two-body law as follows equation:
\begin{equation}\label{eq:K}
   \frac{K}{\rm km\,s^{-1}}= 212.6\, \big(\frac{M_1}{\Msun}\big)^{\frac{1}{3}} \big(\frac{P}{\rm d}\big)^{-\frac{1}{3}} \, q\, (1+q)^{-\frac{2}{3}}  \frac{\sin i}{\sqrt{1-e^2}},
\end{equation}
where $i$ and $e$ are orbital orientation and eccentricity, respectively. The basic distributions of parameters are assumed as follows forms:

\begin{itemize}
  \item Most $e$ are close to $0$. We adopt circular orbit $e=0$ to simplify our calculation.
  \item The distribution of $P$ follows \cite{Raghavan2010}. A log-normal distribution is used to draw random mock $P$ in our method.
  \item The distributions of $i$ and $q$ follow uniform functions.
\end{itemize}

\begin{figure}
	\includegraphics[width=\columnwidth]{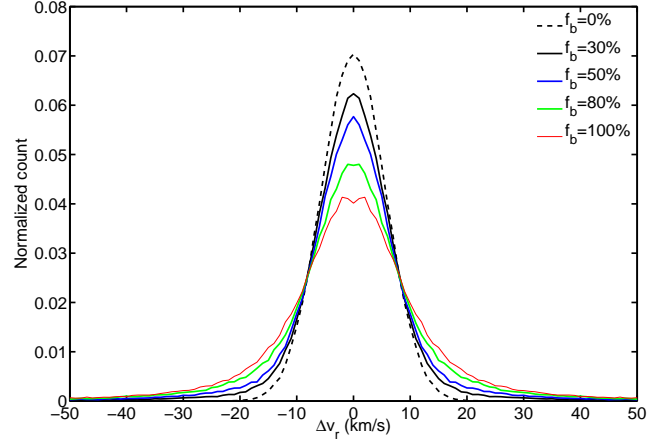}
    \caption{The profile case of $\dv$ derived by our model with different $\fb$. Five curves represent five different input values of $\fb$: $0$, $30$, $50$, $80$, and $100$ per cent, respectively. The $M_1$ is equal to $1\,\Msun$. The x- and y-axis is $\dv$ and normalised count. These profiles don't consider observed errors.}
    \label{fig:model}
\end{figure}

The models with different $\fb$ are shown as Figure \ref{fig:model}. The mock $\dv$ derived by our model depends on $M_1$, which is determined based on $\teff$ and $\feh$ in sample by theoretical stellar isochrones \citep{Girardi2000}. In Section \ref{sec:2d}, the $\teff$ and $\feh$ of each box is used to estimate the $M_1$ for mock binary model. In the other words, each box has a set of model with different $M_1$ and corresponding mock $\dv$. 

Considering orbital phase of two observed epochs and observed random error, we have
\begin{equation}\label{eq:dvr}
  \dv^{(b)} = v_{1}^{(b)}-v_{2}^{(b)} \sim N\big( K (\cos{\phi_1}-\cos{\phi_2}), \sqrt{2}\sig \big),
\end{equation}
which $K$ is the amplitude of RV variation, and $\phi$ is the phase of each observation time. The $\dv$ due to binary stars follows a Gaussian profile with average of $K (\cos{\phi_1}-\cos{\phi_2})$ and standard deviation of $\sqrt{2}\sig$.

If a profile of $\dv$ can be explained by a Gaussian distribution with random error $\sqrt{2}\sig$, this profile is considered to be full of single stars. If $\dv^{(b)}$ can depict a profile of $\dv$, it will be considered to be $\fb$ of $100$ per cent. The optimal combination of $\fb$ and $\sig$ of a sample with given $\dv$ distribution should be fitted at the same time.

\begin{figure}
	\includegraphics[width=\columnwidth]{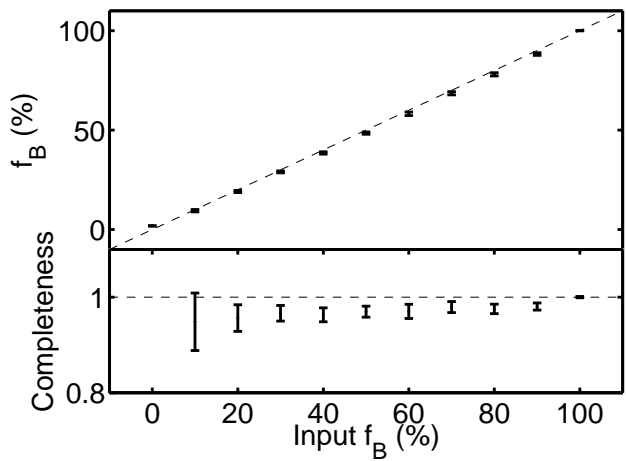}
    \caption{The simulation and test of our method and binary model. \emph{Top}: the comparison between input $\fb$ and derived $\fb$. The dashed line is $1:1$ for reference. Most error bars are smaller than the markers in the figure, so they are not very obvious. \emph{Bottom}: the y-axis is converted into completeness that is defined as the ratio of derived to input $\fb$. The dashed line marks $100\%$ level.}
    \label{fig:sim}
\end{figure}

To test our method, we generate mock samples with known input $\fb$  using model described above and then to simulate the process of $\fb$ estimation. The simulated sample contains 10,000 sources with different given input $\fb$ from $0$ to $100$ per cent. The estimation process of the $\fb$ (and $\sig$) for each input $\fb$ is repeated 100 times to obtain their uncertainties. The $\fb$ of mock samples are estimated as the showing of Figure \ref{fig:sim}. The $M_1$ is $1\,\Msun$ and other orbital parameters adopt sets of model mentioned above. The completeness of estimation of $\fb$ is always larger than $95$ per cent for various input $\fb$.

For real observed sample, we run the process that is similar to above test. The 10,000 mock sources with $100\%$ binary star component are prepared before the fitting process. The normalised profile of $\dv^{(b)}$ is calculated based on mock binary star model. These ``binary stars'' are our binary component whose profile is used to fit the DRV of the real sample. The $\fb$ and $\sig$ are determined by fitting the contribution rates of mock binary star sample and Gaussian error.

\subsection{$\teff$ and $\feh$}
\label{sec:2d}

To deal with our sample in which stellar mass need to be determined before a fitting process of $\fb$. Based on stellar isochrones of MS stars, the $M_1$ is estimated via $\teff$ and $\teff$ in the sample. In our sample, maximum $M_1$ is below to $2\,\Msun$.

\cite{Gao2014} and \cite{Yuan2015} divided field stars into three populations with $\feh$ of the thin disk, thick disk, and halo in the Milky Way. The larger sample allows us to investigate $\fb$ along with $\feh$ in a more detailed way.

Combining above two aims, the sample is divided into $20\times20$ box along with $\teff$ and $\feh$. Considering the errors $(\sim100\,\rm{K},~ \sim0.1\,\rm{dex})$ for $\teff$ and $\feh$ \citep{Gao2015,Ren2016}, the two-dimensional (2D) distribution in $20\times20$ box is smoothed by a $2\times2$ box. In each $(\teff, \feh)$ box, the stellar mass $M_1$ is considered independently when we fit $\fb$ and $\sig$.

Similar to simulation and testing method, fitting process is repeated 100 times to obtain error bars of $\fb$ and $\sig$. If star count is too small, the $\fb$ and $\sig$ fitting cannot output credible results. We don't run the fitting process if the star count in a $(\teff, \feh)$ box is less than 100.

% Sect. 4
\section{Result and Discussion}
\label{sec:result}

\begin{figure*}
	\includegraphics[width=\textwidth]{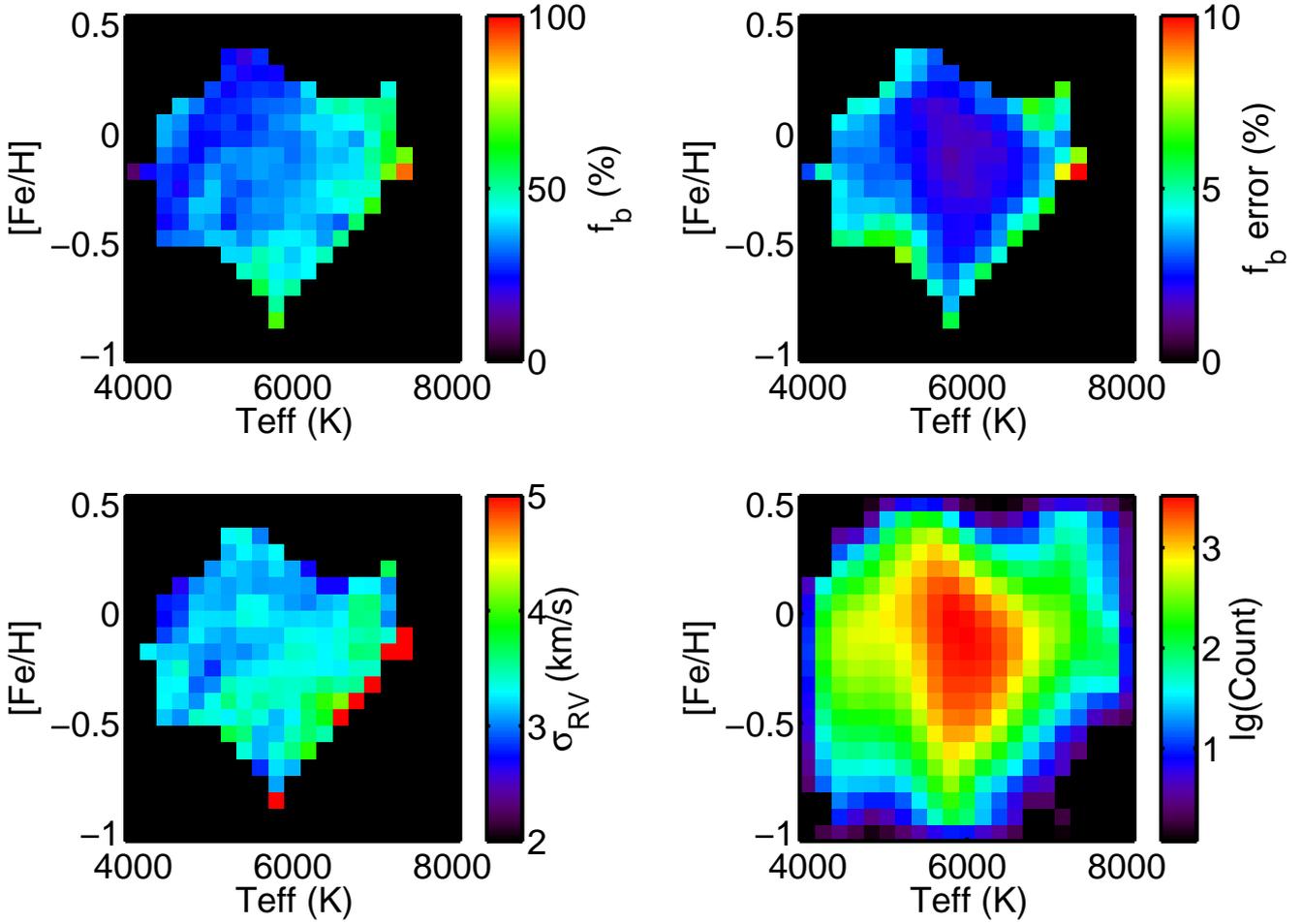}
    \caption{\emph{Upper left}: The distribution of the $\fb$ in the sample. The x-axis is the $\teff$ from $4000$ to $8000\,\rm{K}$ and the y-axis is the $\feh$ from $-1.0$ to $0.5\,\rm{dex}$. The $\fb$, for each $(\teff,\feh)$ sub-sample, is colour-coded by a linear scale from $0$ to $100$ per cent. The black region is a lack of enough data (at least 100 sources). \emph{Upper right}: The distribution of the $\fb$ error in the sample. The x-axis and the y-axis are the same with the left top panel. The $\fb$ error is colour-coded linearly from $0$ to $10$ per cent. \emph{Lower left}: The distribution of the $\sig$. The axis is the same with upper panels. The $\sig$ is colour-coded by a linear scale from $2$ to $5\,\rm{km\,s^{-1}}$. \emph{Lower right}: The distribution of the star count in the sample. The axis is the same with other panels. The star count is colour-coded by a logarithmic scale from $1$ to $3000$.}
    \label{fig:fb}
\end{figure*}

As a showing of the bottom panel in Figure \ref{fig:fb}, edge regions of the colourful part have relative larger $\fb$ error. Especially, the $\fb$ errors at the left and right bottom corners are larger than $10$ per cent. So large uncertainty makes $\fb$ of corresponding $(\teff, \feh)$ boxes short of credibility.

The credibilities of the leftmost and rightmost columns of boxes in Figure \ref{fig:fb} should be doubted due to relative large $\fb$ uncertainty. Except that, the $\fb$ increases along with raised $\teff$. The $\fb$ changes from $\sim 20$ to $\sim 50$ per cent when $\teff$ increases from $4000$ to $7500\,\rm{K}$. The relationship between $\fb$ and $\teff$ won't surprise us because of past observational results \citep[and references therein]{Raghavan2010}. Stars with higher temperature are massive to keep their companions. For metal-rich sample (the upper part of upper left panel of Figure \ref{fig:fb}), the $\fb$ increases from $\sim30$ to $\sim60$ per cent throughout the $\teff$. These results are consistent with previous work \citep{Gao2014,Raghavan2010}. Not only is shown that we have confirmed the relationship between $\fb$ and $\teff$, but also the detailed gradual change of $\fb$ with $\teff$.

%% [Fe/H]
In Figure \ref{fig:fb}, $\fb$ changes in two directions: $\teff$ and $\feh$. The $\fb$ increases when $\feh$ decreases in Figure \ref{fig:fb} even though edge region of colourful boxes have large uncertainty. As showing of Figure \ref{fig:fb}, the $\fb$ is close to $50$ per cent for $\feh<-0.5$ dex. Compared with metal-poor stars, the $\fb$ is only between $20$ and $30$ per cent for near-zero $\feh$. The $\fb$ in halo and thick disk population is almost higher twice than thin disk.

Metal-rich stars with the same mass have slightly lower temperature due to higher opacity. Thus metal-rich stars with the same temperature have a little higher stellar mass. The relationship between $\fb$ and $\feh$ needs to be explained by reasons beyond stellar mass.

%% discussion

Metal rich and K-type stars are most unlikely to have companions. Figure \ref{fig:fb} shows a gradient of $\fb$ from top-left to bottom-right corner. In the area of $4500 \,\rm{K} < \teff < 7000 \,\rm{K}$ and $-0.75\, \rm{dex} < \feh < 0.3\, \rm{dex}$, $\fb$ rises $10$ per cent if $\teff$ increases $\sim 400\, \rm{K}$ and $\feh$ decreases $\sim 0.2 \,\rm{dex}$.

Reasons for the relationship between the $\fb$ and $\feh$ need to be investigated carefully. Why metal abundance affects formation or evolution of binary systems? What roles do time and stellar life span play in this relationship? A few reasons can be listed as follows to explain above trends:

\begin{itemize}
\item The formation of metal-poor binary stars intrinsically are easier than metal-rich binaries.
\item The $\fb$ is uniform in different environments originally, but high $\feh$ leads to complicated interactions of binary systems, which loses companions. 
\end{itemize}

In part of $\teff$ and $\feh$ boxes, small sizes of spectral sample produce large relative uncertainties. To confirm the significant and robust of the results from LAMOST DR3, we need a larger spectral sample with repeat observations, which provides substantial evidence for trends of the $\fb$. 

The distribution of $P$ for $M_1<1.5\,\Msun$ keeps a certain trend \citep{Duchene2013}. For $M_1>1.5\,\Msun$, $\teff>7000\,K$. In each panel of Figure \ref{fig:fb}, these intermediate-mass stars only affect two strips of the rightmost side in the colourful areas. And most influenced boxes have been removed out of our final discussion due to their small size of available sample. It is a remarkable fact that the distribution of $q$ values is assumed to be flat between 0.0 to 1.0, but a more complicated profile of $q$ maybe reduce the ratio of massive secondary stars, which causes smaller $v_r$ shift due to the relative orbital motion. Thus the $\fb$ based on observed $\dv$ could be higher generally when we adopt a non-flat $q$ distribution.

% Sect. 5
\section{Conclusions}
\label{sec:summary}

We deal with a spectral sample containing 365,911 pairs of stellar parameters from repeat observations of LAMOST. The profile of DRV of selected MS stars is used to fit the $\fb$ and $\sig$ in a 2D space of $\teff$ and $\feh$. The fitted two components contain artificial binary sample based on Keplerian orbital motion and Gaussian errors of RV.  

We reveal a 2D trend of $\fb$ with $\teff$ and $\feh$. We suggest that the relationship between $\fb$ and $\teff$ (stellar mass) can be confirmed, but one more factor, $\feh$, affects the $\fb$ in our sample. Metal-poor hot stars are more likely to have a close binary companion than metal-rich cool stars. Meanwhile, metal-rich cool stars have the best RV measurements. The general $\sig$ is about $3.0$ to $3.5\,\rm{km\,s^{-1}}$.

The deeper understanding of the relationship between close binary systems and their $\feh$ requires detailed further study and more spectral data.

\section*{Acknowledgements}
We would like to thank the anonymous reviewer for his/her comments, which significantly improved the paper. Authors are supported by grants No. 11503002 and 11533002 from the National Natural Science Foundation of China (NSFC).  Authors thank helpful discussions with Prof. Biwei Jiang and Dr. Haibo Yuan.


%%%%%%%%%%%%%%%%%%%%%%%%%%%%%%%%%%%%%%%%%%%%%%%%%%

% Don't change these lines
\bsp	% typesetting comment
\label{lastpage}
\end{document}